%% file: 070510LHCpredictarXivb.tex
\documentclass[]{iopart}

\usepackage{graphicx}
\begin{document}

\input{mycommands.tex}

\title{Ratio of charm to bottom \raa as a test of pQCD vs.\ \ads energy loss}

\date{\today}
\vspace{-.15in}
\author{W.\ A.\ Horowitz}
\address{Department of Physics, Columbia University, 538 West 120$^\mathrm{th}$ Street, New York, NY 10027, USA}
\address{Frankfurt Institute for Advanced Studies (FIAS), 60438 Frankfurt am Main, Germany}
\vspace{-.15in}
\begin{abstract}
LHC predictions for the charm and bottom nuclear modification factors, \raacpt and \raabptcomma, using pQCD and \ads drag energy loss models are given.  We show that a new observable, the double ratio $R^{cb}(\eqnpt) = \eqnraacpt / \eqnraabpt$, allows for easy experimental distinction between the two classes of energy loss models.
\end{abstract}
\vspace{-.15in}
The theoretical framework of a \weaklycoupled QGP used in pQCD models that quantitatively describe the \highpt \pizerocomma, $\eta$ suppression at RHIC is challenged by several experimental observables, not limited to \highpt only, suggesting the possibility that a \stronglycoupled picture might be more accurate.  One seeks a measurement that may clearly falsify one or both approaches; heavy quark jet suppression is one possibility.  Strongly-coupled calculations, utilizing the \ads correspondence, have been applied to \highpt jets in three ways \cite{Liu:2006ug,Casalderrey-Solana:2006rq,Herzog:2006gh}
.  We will focus on predictions from the \ads heavy quark drag model 
and compare them to pQCD predictions from WHDG convolved radiative and elastic energy loss and radiative only energy loss \cite{Wicks:2005gt}.
Comparisons between \ads calculations and data are difficult.  First, one must accept the double conjecture of QCD$\leftrightarrow$SYM$\leftrightarrow$\adscomma.  Second, to make contact with experiment, one must make further assumptions to map quantities such as the coupling and temperature in QCD into the SUGRA dual.  For example, the \ads prediction for the heavy quark diffusion coefficient is $D = 4/\sqrt{\lambda}(/2\pi T)$ \cite{Casalderrey-Solana:2006rq}, where $\lambda=g_{SYM}^2 N_c$ is the \thooft coupling.  The ``obvious'' first such mapping \cite{Gubser:2006qh} simply equates constant couplings, $g_s=g_{SYM}$, and temperatures, $T_{SYM}=T_{QCD}$.  Using this prescription with the canonical $N_c=3$ and $\eqnalphas=.3$ yields $D\approx1.2(/2\pi T)$.  It was claimed in \cite{Casalderrey-Solana:2006rq} that $D = 3(/2\pi T)$ agrees better with data; this requires $\eqnalphas\approx.05$.  An ``alternative'' mapping \cite{Gubser:2006qh} equates the quark-antiquark force found on the lattice to that computed using \adscomma, giving $\lambda\approx5.5$, and the QCD and SYM energy densities, yielding $T_{SYM}=T_{QCD}/3^{1/4}$.
The medium density to be created at LHC is unknown
; we will take the PHOBOS extrapolation of $\eqndngslashdy=1750$ and the KLN model of the CGC, $\eqndngslashdy=2900$, as two sample values.
  We will search for general trends associated with \ads drag (denoted hereafter simply as \adscomma) or pQCD as the aforementioned uncertainties mean little constrains the possible normalizations of \ads \raaQ predictions for LHC.

\begin{figure}[htb!]
\begin{center}
$\begin{array}{c@{\hspace{.00in}}c}
\includegraphics[width=.515\columnwidth]{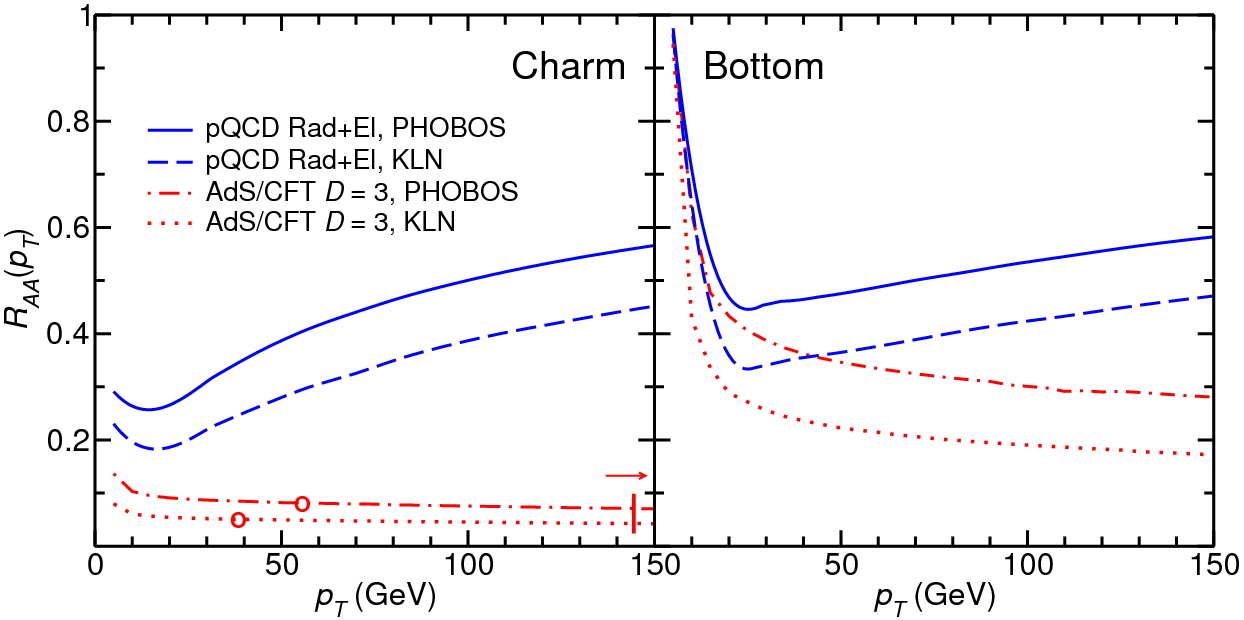} &
\includegraphics[width=.455\columnwidth]{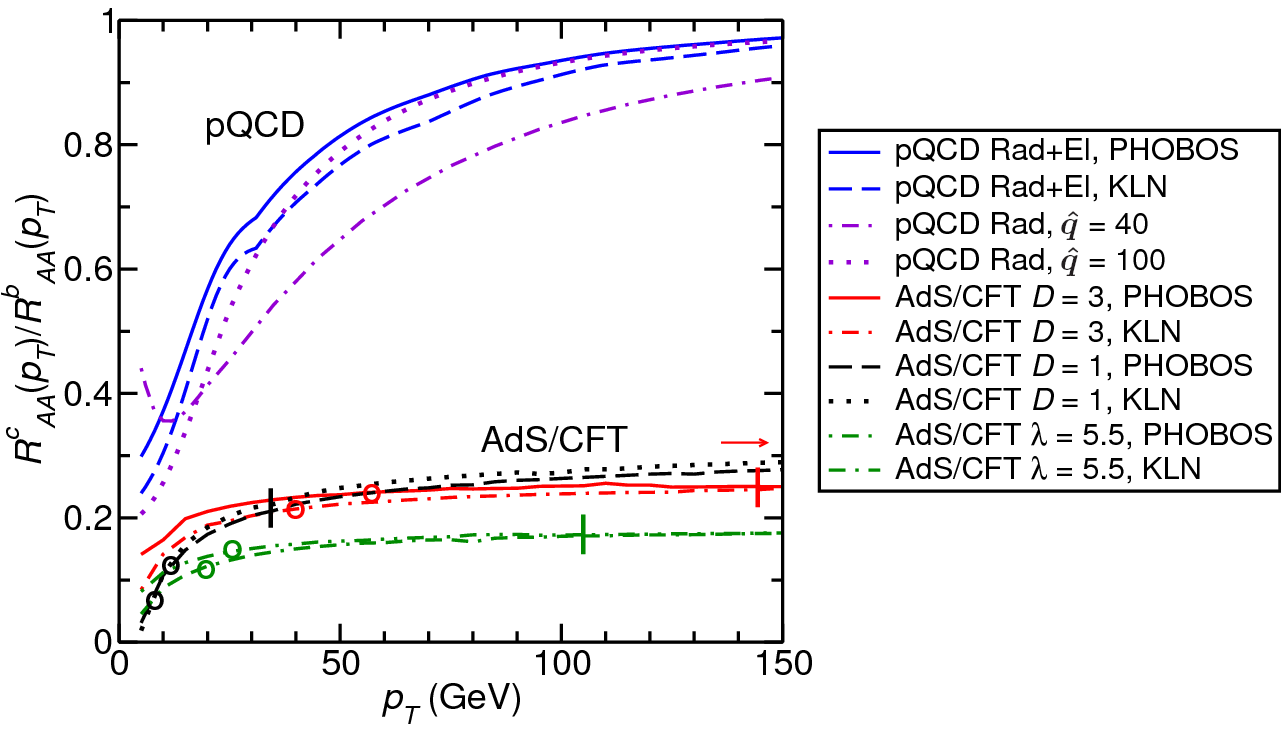} \\[-.1in]
{\mbox {\scriptsize { (a)}}} & {\mbox {\scriptsize { (b)}}}
\end{array}$
\end{center}
\vspace{-.2in}
\caption{(Color Online) (a) Charm and bottom \raapt predictions for LHC.  The generic trend of pQCD curves increasing with \pt while \ads curves decrease is seen for representative input parameters; similar trends occurred for the other input possibilities considered.  (b) Double ratio of charm to bottom \raaptcomma.  pQCD and \ads drag curves fall into two distinct groups; 
the LHC should easily distinguish between the two trends.}
\label{fig1}
\vspace{-.25in}
\end{figure}


The \ads derivation of the drag on a heavy quark yields
$d\eqnpt/dt = -\eqnmuQ\eqnpt = -(\pi\sqrt{\lambda}T_{SYM}^{2}/2\eqnmQ)\eqnpt$ \cite{Herzog:2006gh}, 
giving an average fractional energy loss of $\bar{\epsilon}=1-\exp(-\int \! dt\mu_Q)$.  Asymptotic pQCD energy loss for heavy quarks in a static medium goes as
$\bar{\epsilon}\approx\kappa \eqnalphas L^2 \eqnqhat \log(\eqnpt/\eqnmQ)/\eqnpt$,
where $\kappa$ is a proportionality constant and $L$ is the pathlength traversed by the heavy quark.  Note that \ads fractional momentum loss is independent of momentum while pQCD loss decreases with jet energy.  
The heavy quark production spectrum may be approximated by a slowly varying power law of index $n_Q(\eqnpt)+1$
, then $\eqnraaQ\approx(1-\bar{\epsilon})^{n_Q(\eqnpt)}$.  Since $n_Q(\eqnpt)$ is a slowly increasing function of momentum, we expect $R_{AdS}^Q(\eqnpt)$ to decrease while $R_{pQCD}^Q(\eqnpt)$ to increase as momentum increases.  This behavior is reflected in the full numerical calculations shown in Fig.~1 (a); details of the model 
can be found in \cite{Horowitz:2007su}.  

For large opacity pQCD predicts nearly flat \raaQcomma, masking the difference discussed above.  One can see in Fig.~1 (b) that the separation of AdS/CFT and pQCD predictions is enhanced when the double ratio of charm to bottom nuclear modification, $R^{cb}(\eqnpt)=\eqncbratio$, is considered.  Asymptotic pQCD energy loss goes as $\log(\eqnmQ/\eqnpt)/\eqnpt$, becoming insensitive to quark mass for $\eqnpt\gg\eqnmQ$; hence $R^{cb}_{pQCD}\rightarrow1$.  Expanding the \raa formula for small $\epsilon$ yields $R^{cb}_{pQCD}(\eqnpt)\approx1-p_{cb}/\eqnpt$, where $p_{cb}=\kappa \eqnalphas n(\eqnpt)L^2 \log(m_b/m_c) \eqnqhat$ and $n_c\approx n_b=n$.  Therefore the ratio approaches unity more slowly for larger suppression.  This behavior is reflected in the full numerical results for the moderately quenched pQCD curves, but is violated by the highly oversuppressed $\eqnqhat=100$ curve.  The \ads drag, however, is independent of \ptcomma.  
Approximating the medium with a static plasma of thickness $L$ gives $\eqnraaQ\approx\int_0^L\! d\ell \exp(-n_Q\eqnmuq\ell)\approx1/n_Q\eqnmuq L$ which yields $R^{cb}(\eqnpt)\approx n_b(\eqnpt)m_c/n_c(\eqnpt)m_b\approx m_c/m_b\approx.27$.
This behavior is also reflected in the full numerical results shown in Fig.~1 (b), and so, remarkably, the pQCD and \ads curves fall into easily distinguishable groups, robust to changes in input parameters.
An estimate for the momentum after which corrections to the above \ads drag formula are needed, $\gamma>\gamma_c$, found in the static string geometry is $\gamma_c=1/1+(2\eqnmQ/T\sqrt{\lambda})$ \cite{Gubser:2006nz}.  Since temperature is not constant we show the smallest speed limit, using $T(\tau_0,\vec{x}=\vec{0})$, and largest, from $T_c$, represented by ``O'' and ``$|$,'' respectively.  A deviation of $R^{cb}$ away from unity at LHC in year 1 would pose a serious challenge to the usual pQCD paradigm.  An observation of a significant increase in $R^{cb}$ with jet momenta would imply that the current \ads picture is only applicable at low momenta, if at all.  For a definitive statement to be made a $p+Pb$ control run will be crucial.


\end{document}

%% file: mycommands.tex
\newcommand{\be}{\begin{equation}}
\newcommand{\ee}{\end{equation}}
\newcommand{\bfig}{\begin{figure}}
\newcommand{\efig}{\end{figure}}
\newcommand{\bea}{\begin{eqnarray}}
\newcommand{\eea}{\end{eqnarray}}
\newcommand{\infinitessimal}{\mathrm{d}}
\newcommand{\infinitesmal}{\mathrm{d}}
\newcommand{\infinitesimal}{\mathrm{d}}
\newcommand{\intd}{\mathrm{d}}


\newcommand{\raa}{$R_{AA}$ }
\newcommand{\raacomma}{$R_{AA}$}
\newcommand{\raaphi}{$R_{AA}(\phi)$ }
\newcommand{\raaphicomma}{$R_{AA}(\phi)$}
\newcommand{\raaphipt}{$R_{AA}(\phi;\,\eqnpt)$ }
\newcommand{\raaphiptcomma}{$R_{AA}(\phi;\,\eqnpt)$} 
\newcommand{\raapt}{$R_{AA}(\eqnpt)$ } 
\newcommand{\raaptcomma}{$R_{AA}(\eqnpt)$} 
\newcommand{\eqnraapt}{R_{AA}(\eqnpt)} 
\newcommand{\raaq}{$R_{AA}^Q$ }
\newcommand{\raaqcomma}{$R_{AA}^Q$}
\newcommand{\eqnraaq}{R_{AA}^Q}
\newcommand{\raaqphi}{$R_{AA}^Q(\phi)$ }
\newcommand{\raaqphicomma}{$R_{AA}^Q(\phi)$} 
\newcommand{\eqnraaqphi}{R_{AA}^Q(\phi)}
\newcommand{\raaqphipt}{$R_{AA}^Q(\phi;\,\eqnpt)$ }
\newcommand{\raaqphiptcomma}{$R_{AA}^Q(\phi;\,\eqnpt)$} 
\newcommand{\eqnraaqphipt}{R_{AA}^Q(\phi;\,\eqnpt)} 
\newcommand{\raaqpt}{$R_{AA}^Q(\eqnpt)$ } 
\newcommand{\raaqptcomma}{$R_{AA}^Q(\eqnpt)$} 
\newcommand{\eqnraaqpt}{R_{AA}^Q(\eqnpt)}
\newcommand{\mq}{$m_Q$ }
\newcommand{\mqcomma}{$m_Q$}
\newcommand{\eqnmq}{m_Q}
\newcommand{\muq}{$\mu_Q$ }
\newcommand{\muqcomma}{$\mu_Q$}
\newcommand{\eqnmuq}{\mu_Q}
 
\newcommand{\raaQ}{$R_{AA}^Q$ }
\newcommand{\raaQcomma}{$R_{AA}^Q$}
\newcommand{\eqnraaQ}{R_{AA}^Q}
\newcommand{\raaQphi}{$R_{AA}^Q(\phi)$ }
\newcommand{\raaQphicomma}{$R_{AA}^Q(\phi)$} 
\newcommand{\eqnraaQphi}{R_{AA}^Q(\phi)}
\newcommand{\raaQphipt}{$R_{AA}^Q(\phi;\,\eqnpt)$ }
\newcommand{\raaQphiptcomma}{$R_{AA}^Q(\phi;\,\eqnpt)$} 
\newcommand{\eqnraaQphipt}{R_{AA}^Q(\phi;\,\eqnpt)} 
\newcommand{\raaQpt}{$R_{AA}^Q(\eqnpt)$ } 
\newcommand{\raaQptcomma}{$R_{AA}^Q(\eqnpt)$} 
\newcommand{\eqnraaQpt}{R_{AA}^Q(\eqnpt)}
\newcommand{\mQ}{$m_Q$ }
\newcommand{\mQcomma}{$m_Q$}
\newcommand{\eqnmQ}{m_Q}
\newcommand{\muQ}{$\mu_Q$ }
\newcommand{\muQcomma}{$\mu_Q$}
\newcommand{\eqnmuQ}{\mu_Q}

\newcommand{\raac}{$R_{AA}^c$ }
\newcommand{\raaccomma}{$R_{AA}^c$}
\newcommand{\eqnraac}{R_{AA}^c}
\newcommand{\raacphi}{$R_{AA}^c(\phi)$ }
\newcommand{\raacphicomma}{$R_{AA}^c(\phi)$} 
\newcommand{\eqnraacphi}{R_{AA}^c(\phi)}
\newcommand{\raacphipt}{$R_{AA}^c(\phi;\,\eqnpt)$ }
\newcommand{\raacphiptcomma}{$R_{AA}^c(\phi;\,\eqnpt)$} 
\newcommand{\eqnraacphipt}{R_{AA}^c(\phi;\,\eqnpt)} 
\newcommand{\raacpt}{$R_{AA}^c(\eqnpt)$ } 
\newcommand{\raacptcomma}{$R_{AA}^c(\eqnpt)$} 
\newcommand{\eqnraacpt}{R_{AA}^c(\eqnpt)} 

\newcommand{\raab}{$R_{AA}^b$ }
\newcommand{\raabcomma}{$R_{AA}^b$}
\newcommand{\eqnraab}{R_{AA}^b}
\newcommand{\raabphi}{$R_{AA}^b(\phi)$ }
\newcommand{\raabphicomma}{$R_{AA}^b(\phi)$} 
\newcommand{\eqnraabphi}{R_{AA}^b(\phi)}
\newcommand{\raabphipt}{$R_{AA}^b(\phi;\,\eqnpt)$ }
\newcommand{\raabphiptcomma}{$R_{AA}^b(\phi;\,\eqnpt)$} 
\newcommand{\eqnraabphipt}{R_{AA}^b(\phi;\,\eqnpt)} 
\newcommand{\raabpt}{$R_{AA}^b(\eqnpt)$ } 
\newcommand{\raabptcomma}{$R_{AA}^b(\eqnpt)$} 
\newcommand{\eqnraabpt}{R_{AA}^b(\eqnpt)}

\newcommand{\cbratio}{$\eqnraacpt/\eqnraabpt$ }
\newcommand{\cbratiocomma}{$\eqnraacpt/\eqnraabpt$}
\newcommand{\eqncbratio}{\eqnraacpt/\eqnraabpt}

\newcommand{\raag}{$R_{AA}^g$ }
\newcommand{\raagcomma}{$R_{AA}^g$}
\newcommand{\eqnraag}{R_{AA}^g}
\newcommand{\raagphi}{$R_{AA}^g(\phi)$ }
\newcommand{\raagphicomma}{$R_{AA}^g(\phi)$} 
\newcommand{\eqnraagphi}{R_{AA}^g(\phi)}
\newcommand{\raagphipt}{$R_{AA}^g(\phi;\,\eqnpt)$ }
\newcommand{\raagphiptcomma}{$R_{AA}^g(\phi;\,\eqnpt)$} 

\newcommand{\eqnraagphipt}{R_{AA}^g(\phi;\,\eqnpt)} 
\newcommand{\raagpt}{$R_{AA}^g(\eqnpt)$ } 
\newcommand{\raagptcomma}{$R_{AA}^g(\eqnpt)$} 
\newcommand{\eqnraagpt}{R_{AA}^g(\eqnpt)} 

\newcommand{\RAA}{\raa}
\newcommand{\RAAcomma}{\raacomma}
\newcommand{\RAAphi}{\raaphi}
\newcommand{\RAAphicomma}{\raaphicomma}
\newcommand{\RAAphipt}{\raaphipt}
\newcommand{\RAAphiptcomma}{\raaphiptcomma}
\newcommand{\raapi}{$R_{AA}^\pi$ }
\newcommand{\raae}{$R_{AA}^{e^-}$ }
\newcommand{\raapicomma}{$R_{AA}^\pi$}
\newcommand{\raaecomma}{$R_{AA}^{e^-}$}
\newcommand{\eqnraapi}{R_{AA}^\pi}
\newcommand{\eqnraae}{R_{AA}^{e^-}}

\newcommand{\vtwo}{$v_2$ }
\newcommand{\vtwocomma}{$v_2$}
\newcommand{\vtwopt}{$v_2(\eqnpt)$ }
\newcommand{\vtwoptcomma}{$v_2(\eqnpt)$}
\newcommand{\eqnraa}{R_{AA}}
\newcommand{\eqnraaphi}{R_{AA}(\phi)}
\newcommand{\eqnraaphipt}{R_{AA}(\phi;\,\eqnpt)} 
\newcommand{\eqnRAA}{\eqnraa}
\newcommand{\eqnRAAphi}{\eqnraaphi}
\newcommand{\eqnRAAphipt}{\eqnraaphipt}
\newcommand{\eqnvtwo}{v_2}
\newcommand{\vtwovsraa}{\vtwo vs.~\raa}
\newcommand{\vtwovsraacomma}{\vtwo vs.~\raacomma}
\newcommand{\eqnvtwopt}{v_2(\eqnpt)}

\newcommand{\pp}{$p+p$ }
\newcommand{\ppcomma}{$p+p$}
\newcommand{\dau}{$d+Au$ }
\newcommand{\daucomma}{$d+Au$}
\newcommand{\auau}{$Au+Au$ }
\newcommand{\auaucomma}{$Au+Au$}
\newcommand{\aplusa}{$A+A$ }
\newcommand{\aplusacomma}{$A+A$}
\newcommand{\cucu}{$Cu+Cu$ }
\newcommand{\cucucomma}{$Cu+Cu$}

\newcommand{\rhopart}{$\rho_{\textrm{\footnotesize{part}}}$ }
\newcommand{\rhopartcomma}{$\rho_{\textrm{\footnotesize{part}}}$}
\newcommand{\eqnrhopart}{\rho_{\textrm{\footnotesize{part}}}}
\newcommand{\npart}{$N_{\textrm{\footnotesize{part}}}$ }
\newcommand{\npartcomma}{$N__{\textrm{\footnotesize{part}}}$}
\newcommand{\eqnnpart}{N_{\textrm{\footnotesize{part}}}}
\newcommand{\taa}{$T_{AA}$ }
\newcommand{\taacomma}{$T_{AA}$}
\newcommand{\eqntaa}{T_{AA}}
\newcommand{\rhocoll}{$\rho_{\textrm{\footnotesize{coll}}}$ }
\newcommand{\rhocollcomma}{$\rho_{\textrm{\footnotesize{coll}}}$}
\newcommand{\eqnrhocoll}{\rho_{\textrm{\footnotesize{coll}}}}
\newcommand{\ncoll}{$N_{\textrm{\footnotesize{coll}}}$ }
\newcommand{\ncollcomma}{$N_{\textrm{\footnotesize{coll}}}$}
\newcommand{\eqnncoll}{N_{\textrm{\footnotesize{coll}}}}
\newcommand{\dndy}{$\frac{dN_g}{dy}$ }
\newcommand{\dndycomma}{$\frac{dN_g}{dy}$}
\newcommand{\eqndndy}{\frac{dN_g}{dy}}
\newcommand{\eqndndyabs}{\frac{dN_g^{abs}}{dy}}
\newcommand{\eqndndyrad}{\frac{dN_g^{rad}}{dy}}
\newcommand{\dnslashdy}{$dN_g/dy$ }
\newcommand{\dnslashdycomma}{$dN_g/dy$}
\newcommand{\eqndnslashdy}{dN_g/dy}
\newcommand{\dngdy}{$\frac{dN_g}{dy}$ }
\newcommand{\dngdycomma}{$\frac{dN_g}{dy}$}
\newcommand{\eqndngdy}{\frac{dN_g}{dy}}
\newcommand{\eqndngdyabs}{\frac{dN_g^{abs}}{dy}}
\newcommand{\eqndngdyrad}{\frac{dN_g^{rad}}{dy}}
\newcommand{\dngslashdy}{$dN_g/dy$ }
\newcommand{\dngslashdycomma}{$dN_g/dy$}
\newcommand{\eqndngslashdy}{dN_g/dy}
\newcommand{\as}{\alpha_s}
\newcommand{\alphas}{$\as$ }
\newcommand{\alphascomma}{$\as$}
\newcommand{\eqnalphas}{\as}

\renewcommand{\pt}{$p_T$ }
\newcommand{\pT}{\pt}
\newcommand{\ptcomma}{$p_T$}
\newcommand{\pTcomma}{\ptcomma}
\newcommand{\eqnpt}{p_T}
\newcommand{\ptf}{$p_T^f$ }
\newcommand{\ptfcomma}{$p_T^f$}
\newcommand{\eqnptf}{p_T^f}
\newcommand{\pti}{$p_T^i$ }
\newcommand{\pticomma}{$p_T^i$}
\newcommand{\eqnpti}{p_T^i}
\newcommand{\lowpt}{low-\pt}
\newcommand{\lowptcomma}{low-\ptcomma}
\newcommand{\midpt}{mid-\pt}
\newcommand{\midptcomma}{mid-\ptcomma}
\newcommand{\intermediatept}{intermediate-\pt}
\newcommand{\intermediateptcomma}{intermediate-\ptcomma}
\newcommand{\highpt}{high-\pt}
\newcommand{\highptcomma}{high-\ptcomma}
\newcommand{\Aperp}{$A_\perp$ }
\newcommand{\Aperpcomma}{$A_\perp$}
\newcommand{\eqnAperp}{A_\perp}
\newcommand{\rperp}{$r_\perp$ }
\newcommand{\rperpcomma}{$r_\perp$}
\newcommand{\eqnrperp}{r_\perp}
\newcommand{\eqnrperpHS}{r_{\perp,HS}}
\newcommand{\eqnrperpWS}{r_{\perp,WS}}
\newcommand{\Rperp}{$R_\perp$ }
\newcommand{\Rperpcomma}{$R_\perp$}
\newcommand{\eqnRperp}{R_\perp}

\newcommand{\pizero}{$\pi^0$ }
\newcommand{\eqnpizero}{\pi^0}
\newcommand{\pizerocomma}{$\pi^0$}

\newcommand{\qhat}{$\hat{q}$ }
\newcommand{\qhatcomma}{$\hat{q}$}
\newcommand{\eqnqhat}{\hat{q}}

\newcommand{\gym}{$g_{SYM}$ }
\newcommand{\gymcomma}{$g_{SYM}$}
\newcommand{\eqngym}{g_{SYM}}
\newcommand{\gsym}{\gym}
\newcommand{\gsymcomma}{\gymcomma}
\newcommand{\eqngsym}{\eqngym}
\newcommand{\gs}{$g_{s}$ }
\newcommand{\gscomma}{$g_{s}$}
\newcommand{\eqngs}{g_{s}}
\newcommand{\asym}{$\alpha_{SYM}$ }
\newcommand{\asymcomma}{$\alpha_{SYM}$}
\newcommand{\eqnasym}{\alpha_{SYM}}
\newcommand{\alphasym}{\asym}
\newcommand{\alphasymcomma}{\asymcomma}
\newcommand{\eqnalphasym}{\eqnalsym}

\newcommand{\stronglycoupled}{strongly-coupled }
\newcommand{\weaklycoupled}{weakly-coupled }
\newcommand{\stronglycoupledcomma}{strongly-coupled}
\newcommand{\weaklycoupledcomma}{weakly-coupled}
\newcommand{\ads}{AdS/CFT }
\newcommand{\asd}{\ads}
\newcommand{\adscomma}{AdS/CFT}
\newcommand{\asdcomma}{\adscomma}
\newcommand{\thooft}{'t Hooft }
\newcommand{\thooftcomma}{'t Hooft}

\newcommand{\infinity}{\infty}

\newcommand{\rightleftarrow}{\leftrightarrow}

\newcommand{\eq}[1]{Eq.~(\ref{#1})}
\newcommand{\eqn}[1]{Eq.~(\ref{#1})}
\newcommand{\fig}[1]{Fig.~\ref{#1}}
\newcommand{\figtwo}[2]{Figs.~\ref{#1}, \ref{#2}}
\newcommand{\tab}[1]{Table \ref{#1}}


\newcommand{\captionsize}{\small}